# Study of the properties of Cosmic rays and solar X-Ray Flares by balloon borne experiments

S K Chakrabarti [2,1] *, D Bhowmick[1], S Chakraborty[1], S Palit[1], S K Mondal[3,1], A Bhattacharya[1], S Midya[1], and S Chakrabarti[4,1]

[1] Indian Center for Space Physics, 43 Chalantika, Garia Station Rd., Kolkata 700 084, West Bengal, India
[2] S N Bose National Centre for Basic Sciences, JD Block, Salt Lake, Kolkata, 700 098, West Bengal, India
[3] Sidho-Kanho-Birsa University, Po.& Dt. Purulia, 723 101, West Bengal, India
[4] Department of Physics, Maharaja Manindra Chandra College, Kolkata 700 003, West Bengal, India

*Abstract:* *Indian Centre for Space Physics is engaged in pioneering balloon borne experiments with typical payloads less than ~ 3.5kg. Low cost rubber balloons are used to fly them to a height of about 40km. In a double balloon system, the booster balloon lifts the orbiter balloon to its cruising altitude where data is taken for a longer period of time. In this Paper, we present our first scientific report on the variation of Cosmic Rays and muons with altitude and detection of several solar flares in X-rays between 20keV and 100keV. We found the altitude of the Pfotzer maximum at Tropic of Cancer for cosmic rays and muons and catch several solar flares in hard X-rays. We find that the hard X-ray (> 40keV) sky becomes very transparent above Pfotzer maximum. We find the flare spectrum to have a power-law distribution. From these studies, we infer that valuable scientific research could be carried out in near space using low cost balloon borne experiments.*

## 1. INTRODUCTION

Measurements of high energy radiations and cosmic rays are usually done from space based platforms. About 100 years ago, Victor Hess measured Cosmic rays only up to a height of 5km using a large size balloon. Since then regular large balloon borne experiments are being done. Rocket and satellite based experiments started in 1950s and 1960s and considerable progress has been done to obtain X-ray picture of the Universe. Some of the recent successful, which mainly studied radiations from compact stars and solar X-rays are Rossi X-ray Timing Explorer (RXTE) and Ramaty High Energy Solar Spectroscopic Imager (RHESSI) [1, 2].

Indian Centre for Space Physics (ICSP) is carrying out balloon borne near space experiments at a very low cost [3, 4]. It uses rubber balloons which are otherwise used for meteorological purposes. The basic goal is to have a sustainable programme so that one can obtain quality data from space on a daily basis, which are crucial for space weather studies as well. This is particularly feasible due to rapid miniaturization of high quality instruments with reasonable capacity for photon gathering. In this paper, we present results arising out of our program of last five years in this direction. Instruments are fabricated in house at ICSP. As in large balloon experiments, here too, all payloads are returned to the earth by parachutes after balloon bursts at desirable altitude of about 38-40km. The biggest advantage of using smaller balloons is flexibility in choosing launching site and short time-scale in which an experiment could be prepared. Till date, ICSP has launched 41 Scientific Missions in a series named Dignity. Several other tethered flights are conducted from time to time for test and evaluation of instruments before launch.

## 2. TYPICAL LAUNCH PROFILES

Typically, all balloon launches are carried out from Bolpur or its surroundings. While for smaller payloads a single balloon is used, for heavier payloads and especially for a long duration observation, we use two balloons. Each of these rubber balloons has a mass of about 2000g and can go up to about 37-38km, with a payload of about 2 kg or less. After reaching the burst height, balloon bursts, and equipments return to earth by a parachute. In case a long duration flight is desired, the free lift is adjusted suitably.

Each of the flights has a payload box consisting of main scientific instrument defining main purpose of the Mission. It contains, apart from the parachute, one or two video cameras to monitor balloon bursts and the clouds, GPS tracker units, GPS instruments to give the exact route of flight, accelerometers, gyroscopes and magnetometers. Occasionally, a sun tracker to cross-check if the sun is in the field of view, for solar measurements, is also included. Fig. 1 shows an example of the launch of a double balloon system with a parachute and two payload boxes.

In Table 1, we present a list of Missions whose results have been discussed in the present paper. As mentioned earlier, all payloads are returned by the parachute attached. Our biggest constraint is the total mass of payload, which is a maximum of 3.5kg for a double balloon configuration. However, due to technological advancements and miniaturization, it has become possible to obtain very good scientific data even under these constraints. Because of smaller systems, it takes less than a day to achieve mission readiness.

The launch site of Bolpur (West Bengal, India; 23.66N, 87.7167E) has been so chosen that there are no large lakes/rivers or hills within about 50km radius, making it easier to reach landing sites and recover instruments and the data. Normally there are two seasons when favourable wind conditions prevail at the launch site: In May-June (pre-monsoon) and October-November (post-monsoon) period of each year when winds in different layers roughly cancel out the lateral drifts and as a result payload lands within a short distance from launch site. For a long duration flight, the net path could be several hundred kilometres long. Our location at around the Tropic of Cancer compensates for the Earth's tilt in summer and thus solar data is obtained without any special pointing equipment in summer time.

We can, depending on the mass of payload, use Radio tracking process to continuously track progress and be present at landing site ahead of time, or just use usual GPS tracker to track down landing site. Our GPS tracker regularly measures wind parameters till a height of about 38km. Fig. 2(a) shows components of wind velocity as a function of altitude in four missions, namely, D33, D37, D40 and D41 (marked in the upper panel). Lateral components of velocity are a few tens of meters per second, while the vertical component of velocity is only about 4 m/s. These flights are in pre-monsoon season. For the sake of completeness, we also present in Fig. 2(b) post-monsoon result obtained from Dignity 20 (D20) mission. Note that post-monsoon data at launch site shows the velocity component to have a tendency to be westerly ($V_x > 0$) at all heights.

## 3. ACHIEVEMENT OF LONG DURATION FLIGHTS: LIFT-VALVE EFFECT

Payloads more massive than 2 kg are generally lifted with two balloons. We have developed a unique capability of floating payload for a long period of time (in principle, 5-20 hours) without using any valve to leak out excess gas to stop bursting or dropping ballasts in order to maintain a minimal height as is the practice for larger balloons [5]. This is achieved by a so-called booster-orbiter configuration where lifts of the balloons are adjusted in such a way that only booster bursts at, say, $H_b$ letting orbiter to slowly come down to an equilibrium height of $H_o$ where neutral buoyancy is achieved. Here net lift of orbiter matches payload mass. For this to happen, product $\rho(h)V(h)$ must not be monotonically decreasing function of height (as in a normal spherical Mooney-Rivlin types stretched balloons). Rather, it should be increasing initially to a height above tropopause and then decreases at larger height. Here, $\rho(h)$ and $V(h)$ are the density of the ambient air and the volume of the balloon respectively at a height $h$. Fig. 3 shows product of these functions minus mass of balloon and hydrogen gas (x-axis) as a function of height (y axis) for one of our flights. Since variation of external density with height is different in different latitudes and also can depend on season. Increase in volume with height strongly depends on the material of balloons, and softening techniques controlling the rapture of balloons. Thus, giving a universal analytic expression for the lift is difficult. Due to non-monotonicity of pressure-radius relationship especially when the material is softened [6-8], a typical meteorological balloon shows an interesting property. The curves in Fig. 3 give the lift $L(h)$ at a height $h$. Data of $\rho(h)$ is taken from [9]. In this example, the lifts, $L_o(h)$ for orbiter and $L_b(h)$ for booster, monotonically increase till about $h=15$km where lifts are $L_{vb}$ and $L_{vo}$ respectively. $L_b(h)$ starts to go down till the burst height $H_b$ is reached, as booster bursts and orbiter solely carries payload down. The payload ($M_p=2.4$ kg) was more massive than what the orbiter could lift at burst height ($M_p > L_o(H_b)$). As a result, orbiter balloon's downward motion continues till its lift matched $M_p$. The balloon oscillates around a mean cruising altitude ($H_o$, where, $L_o(H_o) = M_p$) on the lift-curve where $Dh/DL<0$. The so-called lift-valve of orbiter $L_{vo} > M_p$ ensures that such an equilibrium cruising altitude exists. This valve-like property which allows payloads to move up easily, but blocks their downward journey is essential to have long duration flights. The payload can be brought down by ejection system or automatically by cooler atmosphere at night.

Immediately after the burst of a balloon, payload falls almost freely till parachute opens. However, in a booster-orbiter configuration, payload returns more slowly to its cruising altitude.

### 3. ROTATIONAL MOTION OF THE PAYLOAD

It would be instructive to know how directional stability could be achieved in a payload whose mass is at the most 3-3.5kg. In present study, we do not consider cases which require active components to achieve pointing accuracy. Instead, we concentrate on finding instantaneous direction in sky plane at which instrument is pointing. For this, we use magnetometers, accelerometers and gyroscopes in all the flights. Motion of payload has all six components of velocity, i. e., three components of the linear velocity and three components of rotational velocity. None of these components are monotonic and indeed has very peculiar motion which depends on the length of rope, wind condition, mass of payload etc. It is generally noted that the angular motion becomes very slow above ~25km and the rotational period could be as large as 30-60 seconds. Thus, at these heights the instrument could be used as a smooth all sky monitor with a single gear system. For solar observations, our payload can point to the sun during pre-monsoon season without any extra effort. During post-monsoon session, instruments point to the sun only for a fraction of time per rotation. Alternatively, a sun-tracker could be used.

## 5. RESULTS AND DISCUSSIONS

In this paper, we present results of (a) cosmic ray measurements by a Geiger Muller Counter, (b) cosmic ray measurements by both the Hamamatsu made and Bicron made photo multiplier tubes, coupled to a NaI(Tl) crystal and (c) Solar X-ray measurements using Bicron made X-ray detector.

5.1. Cosmic rays by a Geiger counter and other detectors

The payload consists of a Geiger counter (LND712, an end-window type alpha-beta-gamma detector) a biasing voltage supplier and data storage. In Fig. 4, we present the rate at which cosmic rays are detected in Dignity 13 mission at different heights. For the sake of clarity, we present the result with respect to time in minutes (x-axis) since lift-off so that data procured during ascend and descend can be distinguished. We find that Pfotzer maximum [10] at our latitude typically occurs at 16-17 km. The nature of data in both ways is similar, except that time taken between lift-off and burst is much longer than time between burst and landing, which is expected. The count rate between ground and height of Pfotzer maximum is dominated by secondary cosmic rays.

In Fig. 5, we show results of a long duration flight Dignity 26, which has been achieved using two balloons. In this case, a Hamamatsu made Photomultiplier tube (R1548) has been chosen as detector. We show raw data of cosmic rays, along with temperature, pressure and data from accelerometer. Data is obtained for about 12 hours. Accelerometer data (in arbitrary units to compare with other data) shows the burst of booster balloon at around 19000s UT on 23/5/2012. After a transient noise, orbiter cruised for about 9 hours before descending. Landing occurs more than 400 kilometers away at 23.96N, 84.65E (see, Table 1). The orbiter cruised at an altitude of about 25km, though there has been some oscillation around this height. This can be seen from the temperature data also. Temperature data shows higher than expected value during the cruise of the orbiter, possibly because it was directly facing the sun. Cosmic rays and pressure data are consistent with a continuous data acquisition at a height of 25km.

5.2. Muon Detection

For muon detection we use lead shielding which are 1.5cm thick. We put 40.8 mm long and 10mm diameter GM counter inside the lead shield. In Fig. 6, we present the count rate as a function of time in order to show the counts on way up and on way down. Unlike measurements of cosmic rays, we note here that the count rapidly drops to zero above 30km.

5.3. Solar X-ray Flares

Workhorse for the solar X-ray detection has been a 2" diameter Bicron detector with a $40^o$ x $40^o$ degree lead collimator. The collimator and shielding is 0.5mm thick to block 100keV photons. To eliminate charged particles we also put a plastic scintillator (1cm thick) on the top of NaI(Tl) Scintillator. Detector is set to detect in the range of 20keV to 120 keV. We present here a brief report on our efforts to study solar flares in X-rays. .

To enable us to view the sun for a maximum amount of time, we have adjusted tilt of detector with respect to the zenith so that the sun is close to centre of collimator when the balloon is at a high altitude. Sun-sensor stamps every trigger, so as to check whether the sun is inside the collimator. By adjusting the tilt angle, it is also ensured that for every few seconds of solar data, we have the sun outside collimator in order to obtain background. During 2012, when solar activities have been lower, we failed to obtain any major flare, though we have obtained the spectra of quiet sun very well.

As solar activities have increased in 2013, we have detected several solar flares. In Fig. 7, we present light curves (3s average of the raw data) of three solar flares of D33 Mission, which are observed when balloon is at heights of 25km, 28km and 32-34km respectively. Data is compared with GOES (3-25keV) data (green line in online version). General timings of the flare match perfectly. However, since our instrument is not sensitive in 3-20keV at such a low height, we see only the hard components.
Long downward spikes in our data are produced when the sun is not in the field of view of the X-ray detector. At lower altitude, spectrum would have more hard photons due to blockage of softer photons by the remaining atmosphere. As the height of the instrument goes up, more and more soft photons dominate spectrum (as injected solar spectrum itself is dominated by low energy X-rays arising of thermal bremsstrahlung) and observed spectrum should also be close to emitted spectrum [3]. In extreme case, this is precisely why satellite observation is needed to obtain spectrum from ~ 0.1keV and above. Fig. 8 shows channel-energy flux for these flares (whole first flare data and 60s data at rising phase of the other two flares). We find that inner most flare observed at 25km extends to harder X-rays and detected energies of flares progressively become softer and brighter. It is to be noted that at lower energy end of the spectra, counts are sharply reduced. This may be due to extremely high resolution achieved at low temperatures. This is discussed in the context of D40 results where calibrator X-ray lines also found to have a very high resolution.

Dignity 37 mission (D37) has recorded a strong flare which took place on 15[th] May, 2013. At ~ 1:30UT an X-class flare occurred and by the time ICSP launched the balloon at 3:00UT, the flux has become at ~ M1 level. We detected excess X-rays right after the launch from ~ 8-9 km till 30 km. There is no clean data above 30 km and we do not present that here. In Fig. 9, we show count rates (over a ~ 20.3 sq cm. Bicron detector) as a function of height and energy. The 77keV line emitted from our collimator is also shown. As temperature of the front end of detector is decreased, energy resolution of instrument is improved dramatically. Below 60 keV and below Pfotzer maximum (~ 16.5km), we see two components, one from solar flare (higher energy) and the other is from cosmic rays (lower energy). Count rate at ~12-14km, cruising altitudes of commercial aircrafts is found to be important. Judging from what is received at a height of 12km for a C4 flare (which is 25 times weaker than a X1 flare that started at 1:30UT) when our instruments are at that altitude, we can conclude that the commercial flights and defence flights would receive, albeit for a short period, high energy Solar X-rays. The flare is better resolved above ~16.5km, when resolution of the instrument appears to become very good. The resolution at 77 keV is found to be about 4-5% only. This is very intriguing and must be checked by laboratory experiments to ascertain the exact cause. The hard X-ray sky appears to become very transparent above Pfotzer maximum.

In Fig. 10, we compare the energy light curves (energy times count rate) obtained by GOES 15 satellite (un-normalized) in 3-25keV range (green in online version) and D37 light curve obtained in 21-70keV range (red in online version). Below Pfotzer maximum, the light curve is dominated by cosmic rays, though it is a mixture of galactic contribution and solar contribution (as is clear from Fig. 9). Above ~15km, where GOES light curve flattens, solar flare is dominated by high energy photons. At around 20km, the hard photon is not enough to produce a high count as soft photons are blocked. This produces a deep in the light curve. When the payload was at 30km, there are hardly any cosmic ray photons and only the solar contribution is present. The photons in 20-25keV were detected fully. Thus D37 result roughly follows the GOES data at high altitudes.

In Dignity 40 Mission (D40), we have introduced a calibrator to study variation of the resolution with height in more detail. On that day a relatively weak solar flare is detected by GOES-15 satellite when our payload was at an intermediate distance (~ 15-26 km). Data in 26-34km is not recorded. Note the absence of high energy photons (~ 60keV) from 8-9km upward, unlike in D37 mission. This shows that the hard X-ray radiation during a strong flare does cross the Pfotzer maximum and reach the heights of commercial flights. Fig. 11(a) shows 39keV calibrator line and 77keV Pb collimator line as a function of height. The flare becomes very prominent just after Pfotzer maximum. In Fig. 11(b), we show the flare in detail.

In both D37 and D40 results we have corrected for the drift of channels with temperature. However, data has not yet been corrected to subtract the background (less than two counts per second).

In Fig. 12, we present a comparison of the GOES (3-25keV) light curve with that of our D40 mission. We plot the energy flux (un-normalized) in 42-49keV, 49-56keV and 56-63keV energy range. These four curves (marked) are plotted from top most (GOES, red in online version) to bottom (Green, Blue and Violet in online version), respectively. The shapes generally match only above ~ 16.5km, where Pfotzer maximum occurs. This also shows that hard X-ray sky becomes relatively transparent above Pfotzer maximum.

In Fig. 13, we show detected solar spectra in D40 mission at 18.5km (red in online version) and 23km (green in online version). The spectra are obtained with one minute data. Two calibrator lines at 39keV and 77keV are also shown. The spectra fit well in 40-70keV range with power laws having slopes of -4.0 and -4.5 respectively. Surprisingly, the resolutions appear to be very high ( ~ 5%). This is intriguing, as a crystal based detector is not known to have so high resolution [11]. However, we verified that we achieve this resolution again and again under our repeated experiments in nine missions with the same detector. A thorough analysis and laboratory test in space condition is required to ascertain the cause of this behaviour. This may also explain that spectra usually have sharp drop at lower energy side as showed in Fig. 8.

It is instructive to compare the counts anticipated in our detector with counts actually recorded. In our balloon based experiments, in contrast to a satellite based experiment, effects of attenuation of signal due to atmospheric effects (which is highly dependent on the incoming photon energy) is very important. If we consider D37 data at 30km (~ 10 gm/cm$^2$ of atmospheric depth), observed average count rate in 20-25keV is ~ 21 over the whole detector of ~ 20 sq cm (2 inch NaI crystal). This is attenuated by a factor of ~ 300 from the injected photon rate. So, injected photon rate in this range should have been about 300/cm$^2$/s. GOES data of 3-25keV shows energy flux to be 1.3 x 10$^{-4}$ ergs/cm$^2$/s when D37 is at 30km. Assuming a power law distribution $I(E) \sim E^{-2}$ of the solar spectrum, energy content in 20-25keV is about 1/30 of this. Thus the number of injected photons in 20-25 keV should be about ~ 150 /cm$^2$/s, which is in the same ball park of what we detected. We thus show from the light curve and spectral features that our detection is consistent with what is expected of the solar flare inside an atmosphere.

## 6.   CONCLUSIONS

We present results of our consistent and systematic observations by low cost rubber balloons which reach altitudes of ~ 38-40km. We have studied cosmic rays and muons and also detected a large number of X-ray flares of the sun in several of our flights.
With serious constraints over the payload mass, our method is full of innovation. With double balloons we show that a new mathematical concept called lift-valve technique allows us to float payloads by an orbiter balloon for several hours, without usage of normal mechanical valves or ballasts as in usual large balloon flights. We have achieved continuous data acquisition during a 12 hour flight (D26) by this technique. Furthermore, by properly tilting the instruments and judiciously choosing launch time, we can obtain X-ray data from the sun, without using solar tracker.

The biggest advantage of our method is that it is of ultra low cost, each flight being at least a hundred times less expensive than a conventional balloon flight with the same payload. All our payloads are recoverable, making it even more affordable. Our experiments require shorter preparation time and thus, in principle, it is possible to get solar and other high energy data every day from an altitude of near space.

Earth's atmosphere itself is a gigantic detector. It is ionized by X-rays from celestial bodies. Cosmic rays produce showers inside atmosphere. Payloads on their way up by balloons or on their way down by parachutes measure these effects. Thus we not only observe the high energy events, we also study their effects in situ on atmosphere. Effects of the atmosphere on injected spectra are simulated by GEANT4 Monte-Carlo code [3] and results are compared with observations.

So far, ICSP has been carrying out high energy astrophysics related studies. However, measurements of ozone, polluting chemicals, stratospheric cloud compositions, meteorites, aerosols, biological studies are equally possible. In any case, ours is an extraordinary training tool for larger space missions.

One of the exciting results we present was the detection of high energy gamma rays coming from solar events even at the cruising heights of commercial planes. We show that these photons are totally absent in days when there are no major solar events. This is particularly important and requires further study. Similarly intriguing is that just after Pfotzer maximum, resolution appears to improve significantly. One requires to carry out laboratory and numerical studies to ascertain these exciting results.

## ACKNOWLEDGMENTS

We thank the entire balloon team of ICSP, especially, Hiray Roy, Uttam Sardar, Ram Chandra Das and Raj Kumar Maiti who made these experiments successful. We thank Dr. Dipak Debnath and Dr. Ritabrata Sarkar for helpful discussions.

*References*

[1] http://heasarc.gsfc.nasa.gov/docs/xte/XTE.html (Last accessed on 04/04/2012)

[2] http://hesperia.gsfc.nasa.gov/rhessi2/ (Last accessed on 20/06/2012)

[3] S K Chakrabarti et al. *Proceedings of the 20th European Rocket and Balloon Programme and related research* **ESA SP-700** (The Netherlands: ESA Communications) *(ed.)* L. Ouwehand 581 (2011)

[4] S K Chakrabarti et al. *Proceedings of the 21st European Rocket and Balloon Programme and related Research* **ESA SP-701** (The Netherlands: ESA Communications) *(ed.)* L. Ouwehand 663 (2013)

[5] N Yajima, N Izustu, T Imamura and T Abe *Scientific Balloons* (Germany:Springer) (2009)

[6] M Mooney *J. Appl. Phys.* **11** 582 (1940)

[7] R S Rivlin *Phil. Transac. Roy. Soc. London. A, Math. Phys.Sc.,* **241** 379 (1948)

[8] I Muller and H Struchtrup *Math. Mech. Solids* **7** 569 (2002)

[9] A E Hedin *Geophys. Res. Lett.* **96** 1159 (1991)

[10] G A Bazilevskaya and A K Svirzhevskaya *Sp. Sc. Rev.* **85** 431 (1998)

[11] G F Knoll *Radiation Detection and Measurement* (New York: John Wiley) (2010)

| Mission ID | Launch date | Launch site | Major Payload |
|---|---|---|---|
| D13 | 14/5/2011 | Takipur, 23.39N, 87.72E | Geiger Muller Counter |
| D17 | 11/11/2011 | Bolpur, 23.66N, 87.7167E | Muon Detector |
| D19 | 21/11/2011 | | |
| D20 | 22/11/2011 | | X-ray Detector (Hamamatsu PMT & BGO) |
| D26 | 23/05/2012 | Bhaluka, 22.37N, 88.445E | X-ray detector (Hamamatsu PMT & NaI) |
| D27 | 24/05/2012 | | X-ray Detector (Bicron) |
| D28 | 25/05/2012 | | |
| D29 | 04/06/2012 | | |
| D33 | 25/04/2013 | Bolpur, 23.66N, 87.7167E | |
| D37 | 15/05/2013 | | |
| D39 | 17/05/2013 | | |
| D40 | 18/05/2013 | | |
| D41 | 19/05/2013 | | |

*Table 1*

*Table 1. Some Dignity Mission launch parameters and the major instruments which were on board. In Column 1, the Mission Identification number (Dxx stands for Dignity xx) is given. Column 2 is the date of launch of the flight, Column 3 contains the launch site and Column 4 contains the major instrument onboard.*

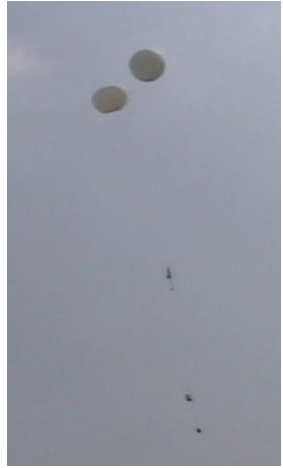

Figure 1

Fig. 1 A double balloon being launched with a single parachute, communication box and the payload.

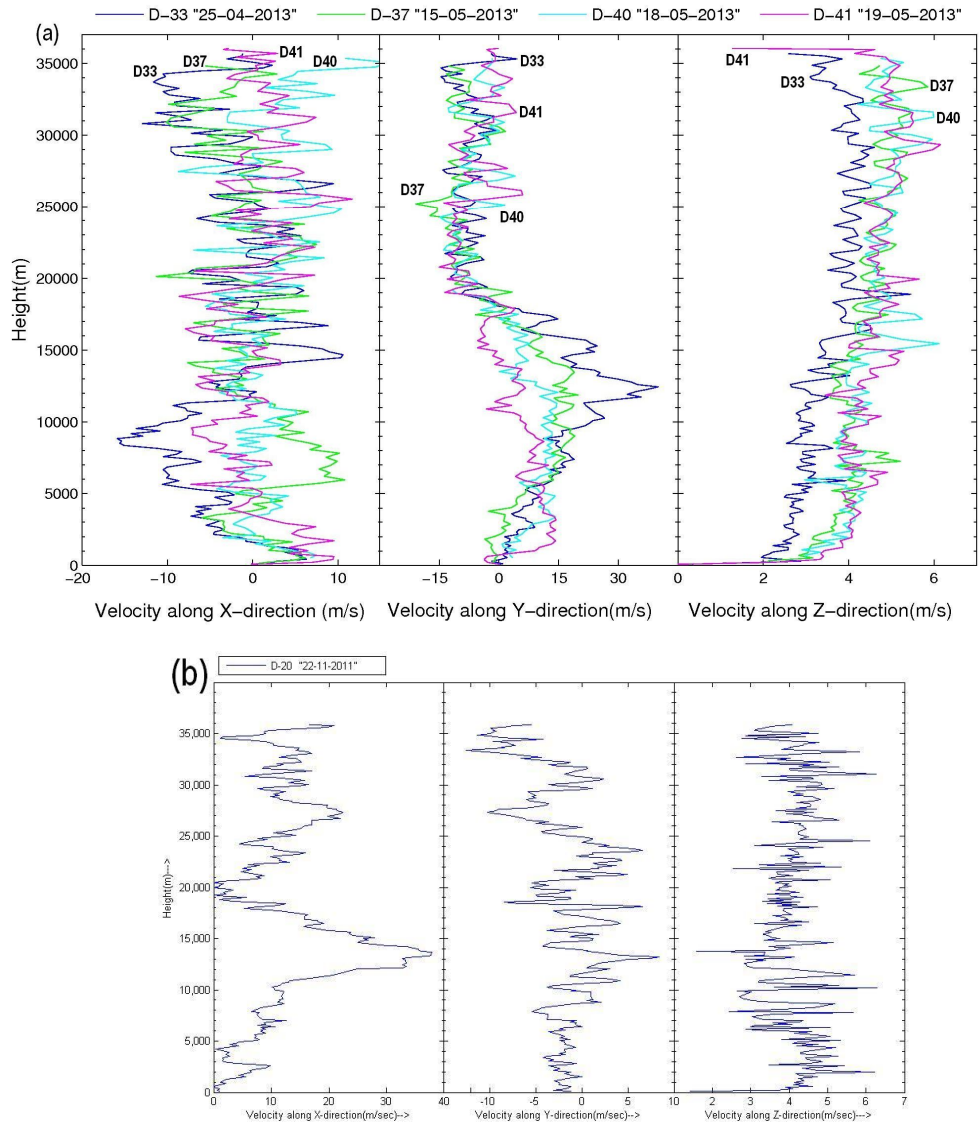

Figure 2(a-b)

Fig. 2(a-b) X, Y, and Z components of wind velocity obtained by us at our latitude in (a) pre-monsoon season in Dignity missions (marked on each curve) 33 (D33), 37 (D37), 40 (D40) and 41 (D41) and (b) in post-monsoon season in Dignity 20 (D20) mission.

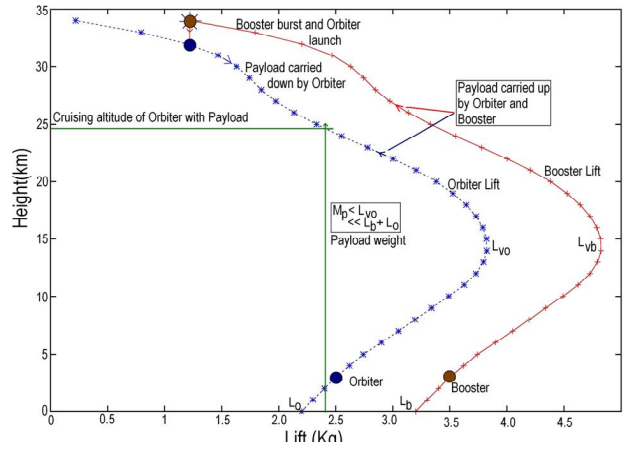

Figure 3

Fig. 3: Working principle of a Booster-Orbiter configuration. Lift-curves of the Booster and the Orbiter are shown. A payload of mass less than the lift-valve of the Orbiter ($M_p < L_{vo}$) can cruise at an equilibrium altitude till it is brought down by ejection or by cold ambience at night.

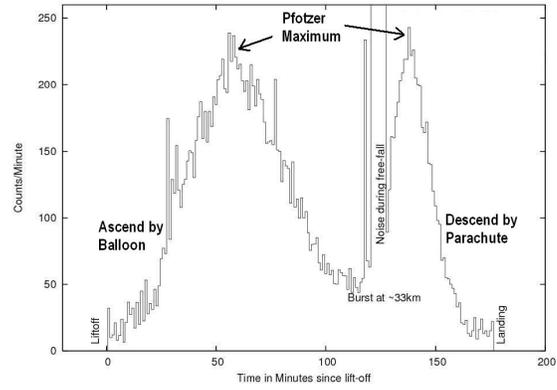

Figure 4

Figure 4: Variation of Cosmic ray count rates while going up and coming down in the Dignity 13 (D13) mission. The Pfotzer maximum is at around 16-17 km.

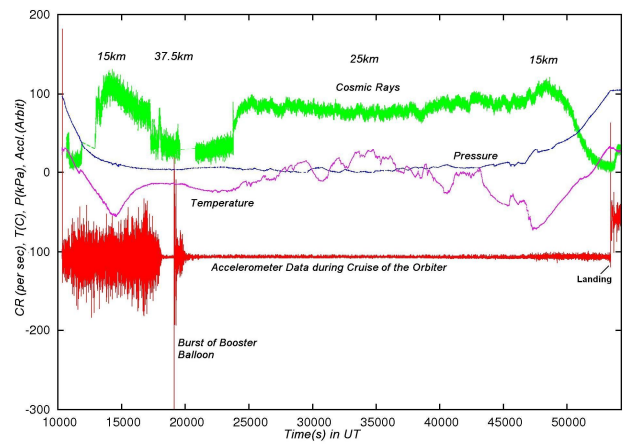

Figure 5

Figure 5. The cosmic rays, pressure, temperature and the Accelerometer data (marked) for a long duration flight (Dignity 26) which lasted for 12 hours. The booster burst at 37.9km and the orbiter was cruising at 25km. There were clear indications (temperature data) of some oscillations of height. Accelerometer data indicates that the flight was smooth.

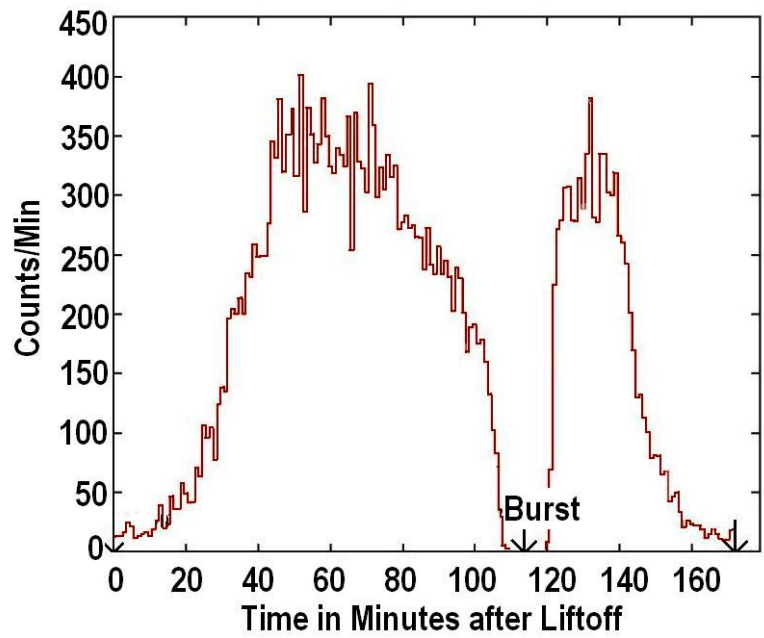

Figure 6

Fig. 6 Muon count rate as a function of time since liftoff. This is the result of Dignity 19 mission. A broad peak occurs at around 15-17 km. There are no muons above ~30km.

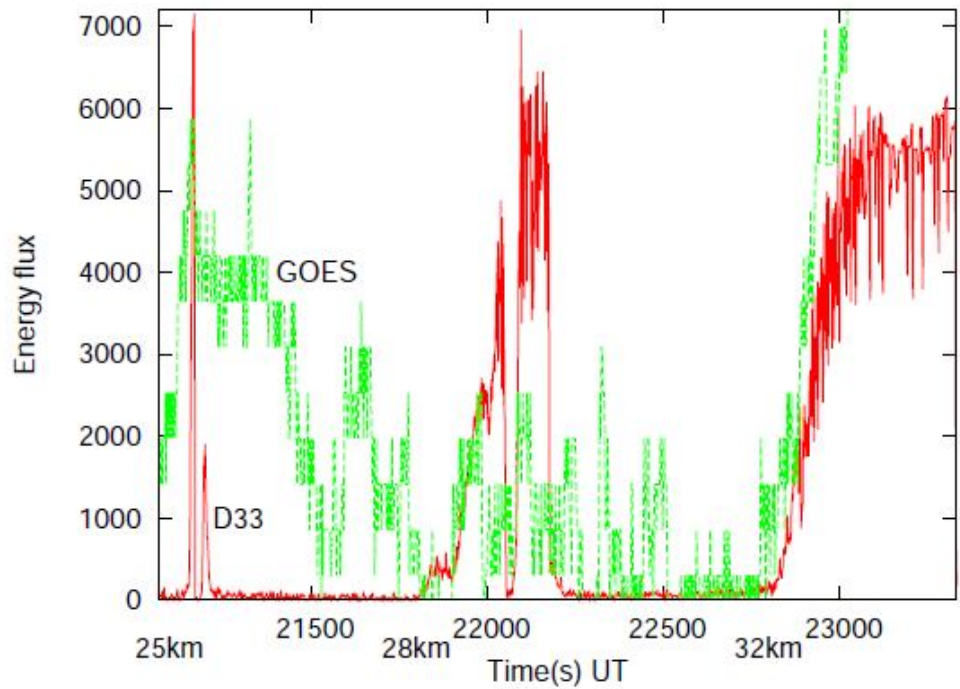

Figure 7

Fig.7 Raw light curves of the three flares observed in D33 mission (25 Apr, 2013; red in online version) as compared with 3-25keV GOES satellite light curve (green in online version). The softer component of 3-20keV was invisible to D33.

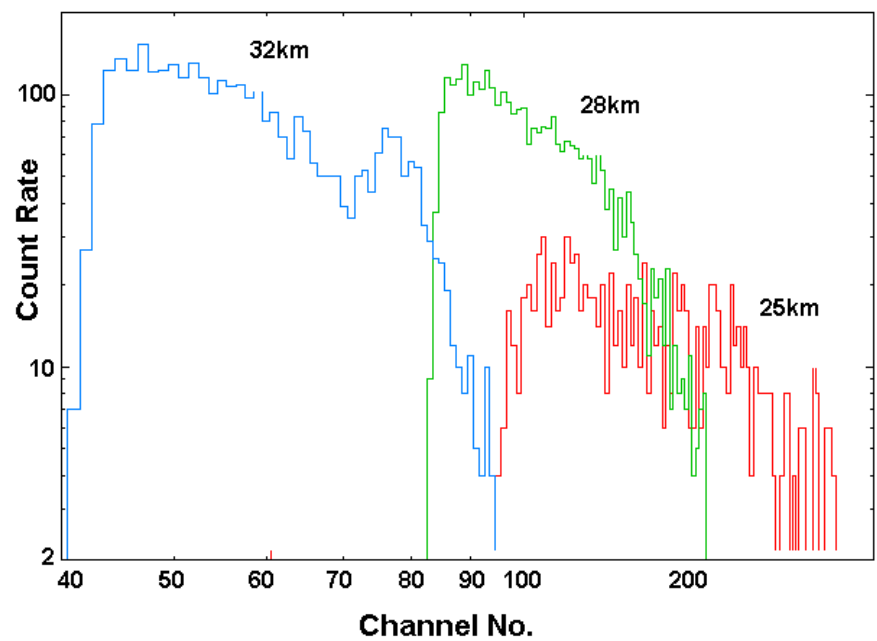

Figure 8

Fig. 8 Channel vs. Photon count rates of the rising phase of the three flares shown in Fig. 7. Altitudes of the payload are marked.

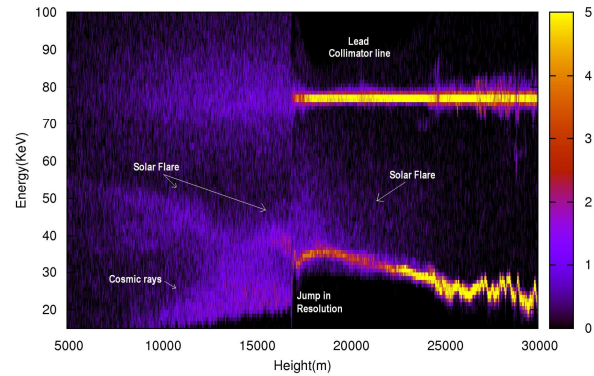

Figure 9

Fig. 9 Dynamical spectra of the spectrum from 5km till 30km. The line at 77keV from the lead collimator is also shown. The energy resolution improved at the temperature is reduced above 17 km, because of which the solar flare could be seen clearly from this height (marked).

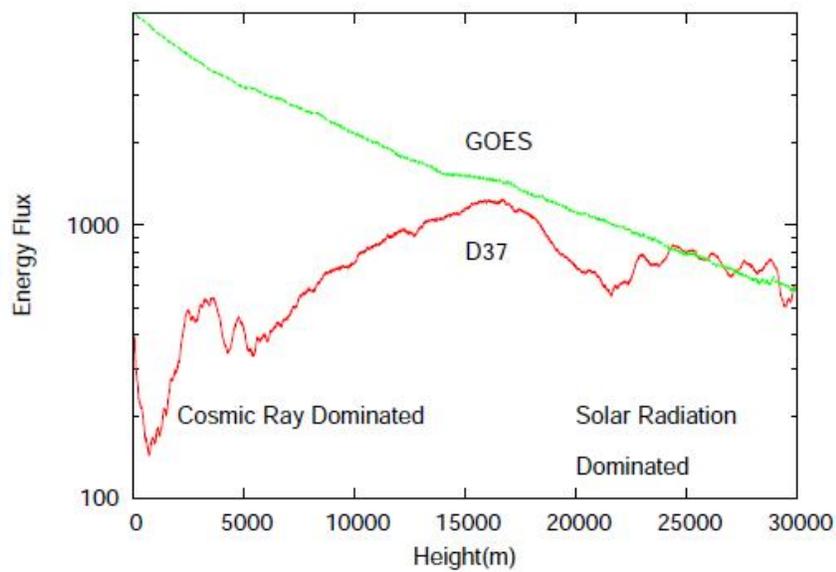

Figure 10

Figure 10: Light curve of 21-70 keV energy flux in D37 averaged over 60s is compared with GOES 15 4-30keV energy light curve. The difference decreases as the height increases. Below 15 km, the energy is dominated both by cosmic rays and solar X-rays. Immediately above 15 km, the Pfotzer maximum, the atmospheric window almost totally opens for hard photons. At around 30km, the shape of D37 light curve generally matches with the GOES light curve.

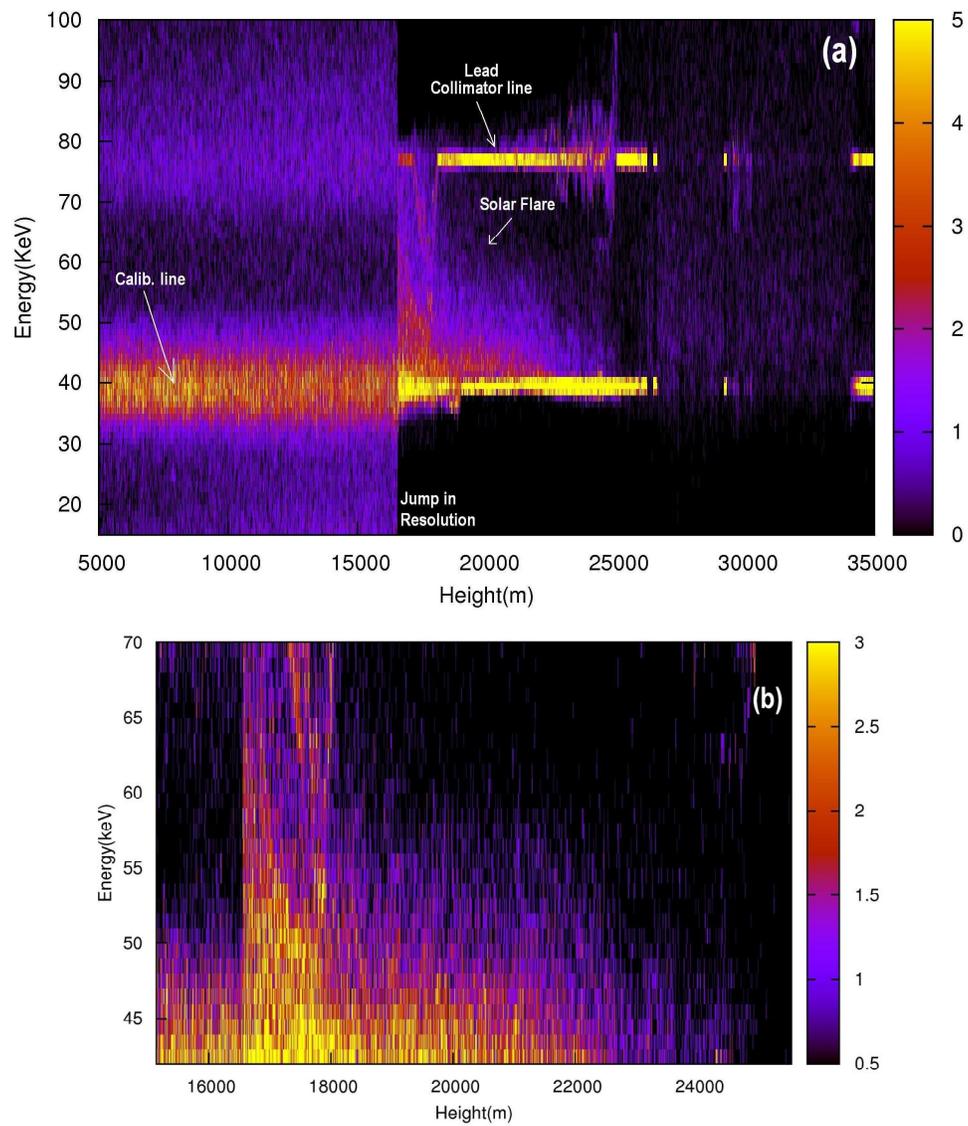

Figure 11

Figure 11: Calibrated dynamical spectrum showing a 39keV and the collimator line 77keV. We see the remnant of a C1 solar flare in (a) clearly after the Pfotzer maximum. In (b), we show the flare in detail between 42 and 70 keV in order to avoid contamination from both the calibrator lines.

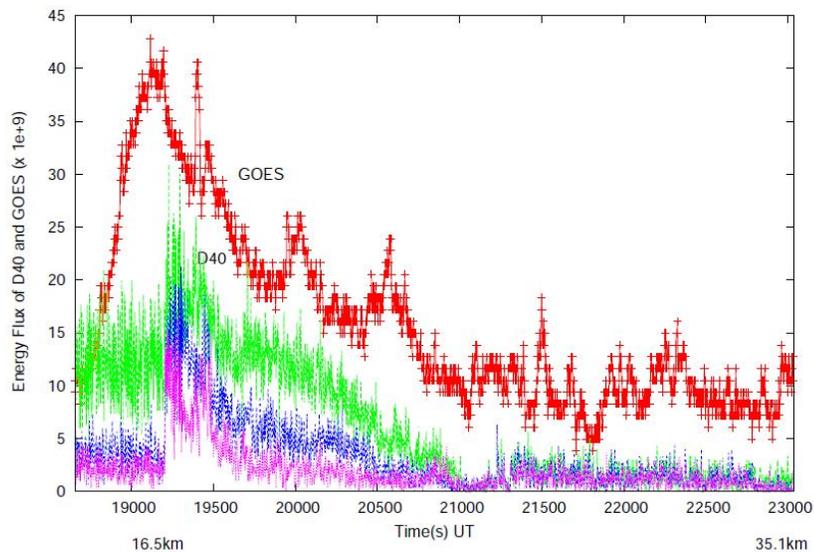

Figure 12

Fig. 12: Energy flux (un-normalized) of the GOES satellite data (Red in online version) of 3-25keV (top most) is compared with D40 data, from top to bottom) in 42-49keV, 49-56keV and 56-63keV energy windows (online version Green, Blue and Violet curves respectively).

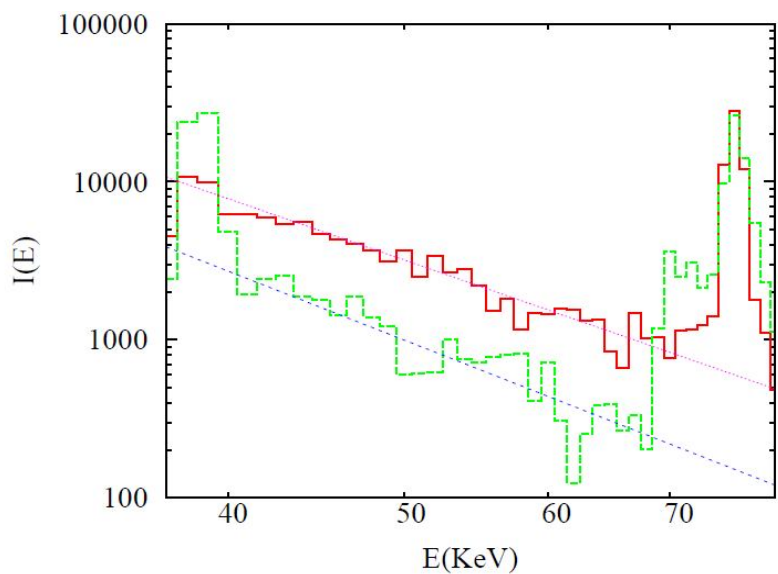

Figure 13

Figure 13: Calibrated solar X-ray energy spectrum as observed from a height of 18.5km (upper curve, red in online version) and 23 km (lower curve, green in online version) with an calibrator line (39keV) and the collimator line (77keV). The spectra fit well in 40-70keV range with power laws having slopes of -4.0 and -4.5 respectively.